\documentclass[conf]{new-aiaa}
\usepackage[utf8]{inputenc}

\usepackage{graphicx}
\usepackage{amsmath}
\usepackage[version=4]{mhchem}
\usepackage{siunitx}
\usepackage{longtable,tabularx}
\usepackage{bm}
\usepackage{physics}

\usepackage{etoolbox}
\makeatletter
\patchcmd{\ttlh@hang}{\parindent\z@}{\parindent\z@\leavevmode}{}{}
\patchcmd{\ttlh@hang}{\noindent}{}{}{}
\makeatother

\newcommand{\mubold}{\bm{\mu}}
\newcommand{\alphabold}{\bm{\alpha}}

\title{Neural-Network-Augmented Projection-Based Model Order Reduction for Mitigating the Kolmogorov Barrier to Reducibility of CFD Models}

\author{Joshua L. Barnett \footnote{Graduate Student, Department of Mechanical Engineering, Durand Building, Room 224, Stanford University, Stanford, 
CA 94305-3035, USA. Email: \texttt{jb0@stanford.edu.}}}
\author{Charbel Farhat\footnote{Vivian Church Hoff Professor of Aircraft Structures, Department of Aeronautics and Astronautics, Department of Mechanical Engineering, and
Institute for Computational and Mathematical Engineering, Durand Building, Room 257, Stanford University, Stanford, CA 94305-3035. AIAA Fellow. Email: \texttt{cfarhat@stanford.edu}.}}
\author{Yvon Maday \footnote{Laboratoire Jacques-Louis Lions (LJLL), Sorbonne Universit\'e and Universit\'e de Paris Cit\'e, CNRS, F-75005 Paris, France. Email: \texttt{yvon.maday@sorbonne-universite.fr}.\\ {\bf This paper has been accepted for publication in 2023 AIAA SciTech Forum - 23–27 January 2023}}}
\affil{Stanford University, CA 94305, USA\\ Sorbonne Universit\'e and Universit\'e de Paris Cit\'e, France}

\begin{document}
\maketitle
\begin{abstract}

Inspired by our previous work on mitigating the Kolmogorov barrier using a quadratic approximation manifold \citep{barnett2022quadratic}, we propose in this paper
a computationally tractable approach for combining a projection-based reduced-order model (PROM) and an artificial neural network (ANN) for mitigating the Kolmogorov
barrier to reducibility of convection-dominated flow problems. The main objective the PROM-ANN concept that we propose is to reduce the dimensionality of the online 
approximation of the solution beyond what is possible using affine and quadratic approximation manifolds. In contrast to previous approaches for constructing arbitrarily 
nonlinear manifold approximations for nonlinear model reduction that exploited one form or another of ANN, the training of the PROM-ANN we propose in this paper does not 
involve data whose dimension scales with that of the high-dimensional model; and this PROM-ANN is hyperreducible using any well-established hyperreduction method. 
Hence, unlike many other ANN-based approaches, the PROM-ANN concept we propose in this paper is practical for large-scale and industry-relevant CFD problems. Its potential
is demonstrated here for a parametric, shock-dominated, benchmark problem.

\end{abstract}


\section{Introduction}

Projection-based model order reduction (PMOR) allows for the parsimonious representation of a parameterized, semi-discrete or discrete high-dimensional model (HDM), through the construction of an 
associated, lower-dimensional, projection-based reduced-order model (PROM). A second-level approximation known as hyperreduction transforms a PROM into a hyperreduced counterpart (HPROM) capable of 
operating in real-time, while preserving the desired level of accuracy. As such, an HPROM is essential for many time-critical applications such as simulation-driven design and optimization, optimal 
control, uncertainty quantification, and digital twinning, to name only a few. 

Traditionally, PMOR has been performed using best-approximations on affine subspaces of dimension $n \ll N$, where $n$ and $N$ denote the dimensions of the PROM/HPROM and underlying HDM, 
respectively.  It has demonstrated significant success for a large number of parametric applications grounded in elliptic, parabolic, and second-order hyperbolic partial differential equations (PDEs) -- 
for example, problems in solid mechanics \cite{zahr2017multilevel, he2020situ}, heat conduction (or diffusion) \cite{rozza2008reduced}, structural dynamics \cite{farhat2014dimensional}, and wave 
propagation \cite{egger2018structure, farhat2020computational}. It has also demonstrated success for parametric, linearized CFD and coupled aeroelastic 
computations \cite{thomas2003three, amsallem2008interpolation}. For nonparametric RANS-based \cite{grimberg2021mesh} and LES-based \cite{grimberg2020stability} CFD computations, PMOR has delivered 
impressive speedup factors including for compressible flow problems past complex three-dimensional geometries \cite{grimberg2021mesh}; and in the presence of shocks \cite{tezaur2022robust}. 
For parametric counterpart flow computations -- and for that matter, for most first-order hyperbolic PDEs describing transport (or advection) problems -- traditional PMOR suffers from the slow decay of 
the Kolmogorov $n$-width. 

The Kolmogorov $n$-width $d_n(\mathcal M)$ describes the {\it worst-case error} with respect to a subset $\mathcal M$ of a normed space $\mathcal S$ arising from a projection onto the 
best-possible linear subspace of $\mathcal S$ of dimension $n$ \cite{greif2019216}, represented here by a right reduced-order basis (ROB) $\mathbf{V} \in \mathbb{R}^{N \times n}$. Hence, 
it sets a limit to any 
PROM-based approximation: for this reason, it is often referred to as a ``barrier'' to reducibiity. In practice, the Kolmogorov $n$-width affects most convection-dominated (flow) models: it
implies a dimension $n$ that is sufficiently large to challenge the construction of a traditional PROM, or simply defeat the purpose of PMOR.

A variety of approaches have been formulated so far to attempt to mitigate the Kolmogorov barrier to PMOR. Typically, these approaches involve altering the traditional affine subspace approximation 
to incorporate
some nonlinearity into it. The simplest of them substitutes the affine subspace approximation with a piecewise counterpart \cite{amsallem2012nonlinear} constructed by partitioning of the
nonlinear high-dimensional solution manifold into regions and computing in each region a local right ROB associated with a local affine subspace approximation. This approach was shown in 2016
to perform in real-time on a laptop, for steady-state RANS computations associated with the NASA Common Research Model \cite{rivers2014experimental} and a parametric
HDM of dimension greater than $N = 68\times 10^6$, in the transonic regime characterized by shocks, at a high Reynolds number ($Re = 5 \times 10^6$), and in a shape parameter domain $\mathcal D$ of 
dimension $N_{\mathcal D} = 4$ \cite{washabaugh2016}. Nevertheless, due to the Kolmogorov barrier, one can reasonably expect such a PMOR method to face increasing difficulties at delivering a similar 
performance for increasing values of $N_{\mathcal D}$.

Registration methods have also been proposed for addressing the Kolmogorov $n$-width issue. From a mathematical viewpoint, these methods transform the traditional affine subspace approximation
into a nonlinear approximation by applying a nonlinear parametrization to the affine subspace 
\cite{welper2017interpolation, cagniart2019model, black2020projection, mojgani2020physics, reiss2021optimization, ferrero2022registration, mirhoseini2022model} -- usually
through a transformation of the given spatial domain. As such, these methods involve the motion and/or deformation of the mesh underlying the semi-discrete or discrete HDM. For this reason, they have 
been limited so far to simple geometries; or are applicable to complex ones but are in this case vulnerable to mesh entanglements -- particularly for large-scale CFD meshes associated with high Reynolds 
number flows. From an intuitive viewpoint, registration methods aim at determining domain mappings that cause parametric features of the solution to align in a reference domain for a set of 
training points sampled in the parameter domain $\mathcal D$, which removes the convective nature of the solution, lowers the Kolmogorov $n$-width, and improves the usefulness of the sampled solution 
snapshots when an unsampled parameter point is queried in $\mathcal D$.

Quadratic approximation manifolds have also been proposed, in two different forms, to mitigate the Kolmogorov barrier: a simulation-free form for structural dynamic applications \cite{jain2017quadratic}; 
and a data-driven form that was demonstrated for convection-dominated turbulent flow problems and other transport problems \cite{barnett2022quadratic}. They have also been proposed for performing 
non-intrusive PMOR using an operator inference method \cite{geelen2023operator}. 

Most relevant to this paper are approaches that have been proposed for constructing arbitrarily nonlinear manifold approximations for PMOR using some form of artificial neural networks (ANNs).
One such approach was introduced in \cite{lee2020model}, in the context of structured meshes, and based on convolutional autoencoders. During training, they map the high-dimensional 
solution to a latent space whose dimension approaches that of the intrinsic solution manifold. Extensions to unstructured meshes were performed in \cite{fresca2021comprehensive}, then in
\cite{gruber2022comparison}, using various means, incuding replacing the convolutional autoencoder with a graph convolutional ANN \cite{gruber2022comparison}. The success of all these approaches however 
comes at a hefty price, as convolutional networks are more expensive to train and evaluate than conventional deep ANNs, due to their greater inter-connectivity that enables their greater expressiveness. 
For this reason, alternative techniques based on conventional ANNs have been more recently proposed for mitigating the Kolmogorov barrier. These include the techniques described in 
\cite{papapicco2022neural, peng2023learning}, which use deep ANNs to learn transformation maps such as those used by registration methods that eliminate the convection nature of the solution, 
without deforming the underlying mesh however.

It should be noted that all ANN-based methods outlined above suffer from the following issues. First, the size of their networks scales with the large dimension $N$ of the HDM. Hence, for 
large-scale computational models such as those associated with flows at high Reynolds numbers and past complex geometries, the training of such networks is computationally intractable. 
This explains why in most if not all cases, these approaches have been demonstrated on simple one-dimensional applications, occasionally on two-dimensional (2D) ones, and usually without
emphasizing offline (training) and/or online (execution) wall clock time. Second, the hyperreduction of PROMs constructed using these approaches remains a challenge, although some 
work in this area has recently appeared in \cite{kim2022fast} for the case of convolutional autoencoders.

The present work is inspired by our previous work on mitigating the Kolmogorov barrier, which focused on a data-driven approach for constructing a quadratic approximation manifold 
\citep{barnett2022quadratic}. It proposes a more general \textit{ANN-augmented} PROM that is referred to here and throughout the remainder of this paper as PROM-ANN. The main objective
of the PROM-ANN is, for a fixed level of accuracy, to reduce the dimensionality of the online approximation of the solution beyond what was possible using the quadratic approximation,
and dramatically beyond what is achievable using the traditional affine subspace approximation. In contrast to the approaches outlined above for constructing arbitrarily nonlinear manifold approximations 
for PMOR using some form of ANN, the training of the PROM-ANN we propose in this paper does \textit{not} involve data whose dimension scales with that of the HDM; and this PROM-ANN \textit{is} 
hyperreducible using any well-established hyperreduction method. Consequently, unlike the aforementioned ANN-based approaches, the PROM-ANN concept we propose in this paper is practical for 
large-scale and industry-relevant problems which, in the context of CFD, tend to have hundreds of millions of degrees of freedom (dofs). 

While the scope of the PROM-ANN concept we propose here is equally applicable to HDMs arising from the semi-discretization or dicretization of ellpitic, parabolic, and both types of
hyperbolic PDEs, we focus in this paper on the case of first-order hyperbolic PDEs, as the PMOR of the corresponding HDMs is most vulnerable to the Kolmogorov barrier. Consequently,
we ground our work in the nonlinear least-squares Petrov-Galerkin (LSPG) PMOR method \cite{carlberg2011efficient, carlberg2013gnat} which represents the state of the art of PMOR for 
convection-dominated turbulent flow problems. However, we note that the appropach we propose in Section \ref{sec:NONLA} for constructing an arbitrarily nonlinear approximation manifold using a 
right ROB and an ANN is equally applicable to Galerkin PMOR methods.

To achieve computational efficiency, we choose to hyperreduce our PROM-ANN using the energy-conserving sampling and weighting (ECSW) \cite{farhat2014dimensional} method. Although originally 
designed in the context of finite element approximations, solid mechanics, and Galerkin PROMs, this hyperreduction method was recently extended in \cite{grimberg2021mesh} to that of finite volume (FV)
approximations, CFD, and LSPG PMOR \cite{grimberg2021mesh}; and was shown to deliver impressive speedup factors while preserving a desirable level of accuracy. Nevertheless, we note that
any other hyperreduction method such as the empirical interpolation method (EIM) \cite{barrault2004empirical} and its discrete version (DEIM) \cite{chaturantabut2010nonlinear} 
can be considered to hyperreduce a PROM-ANN.

The remainder of this paper is organized as follows. In Section \ref{sec:BACK}, we set the scope of this work and overview the LPSG and ECSW methods.
In Section \ref{sec:NONLA}, we present the main underpinning of the PROM-ANN concept -- namely, a computationally tractable approach for constructing an arbitrarily nonlinear approximation manifold 
using a right ROB and a deep ANN. In Section \ref{sec:ADAP}, we adapt LSPG and ECSW to this concept and comment on the additional computational complexity and additional storage requirement 
entailed by the adaptations. In Section \ref{sec:APP}, we apply our proposed PMOR technology to a popular, parametric,
shock-dominated benchmark problem that is vulnerable to the Kolmogorov barrier. We contrast the performance of this computational technology with that of the traditional PMOR method and highlight 
its superiority. Specifically, for the same level of desirable accuracy, we report 50-fold and 6-fold reductions of the wall clock time required by the HDM and a traditional PROM to solve the
benchmark problem, respectively. In Section \ref{sec:CONC}, we offer concluding remarks.

\section{Background}
\label{sec:BACK}

Throughout this paper, we consider the following parametric, nonlinear, semi-discrete, first-order initial problem
\begin{equation} 
	\begin{aligned} 
		\mathbf{M}(\mubold) \mathbf{\dot{u}}\left(t; \mubold\right) + \mathbf{f}\left(\mathbf{u}\left(t; \mubold\right); \mubold\right) - \mathbf{g}\left(t; \mubold\right) & = \mathbf{0} \\ 
		\mathbf{u}\left(0; \mubold\right) & = \mathbf{u}^0(\mubold) 
		\label{eq:hdm} 
	\end{aligned}
\end{equation}
where $t \in (0, T_f]$ denotes time and $(0, T_f]$ denotes the time-interval of interest; $\mubold \in \mathcal{D} \subset \mathbb{R}^{N_{\mathcal D}}$ denotes a vector-valued parameter of dimension 
$N_{\mathcal D}$ and is often referred to as a ``point'' in the parameter domain $\mathcal{D}$; a dot denotes a time-derivative; $\mathbf{u}\in\mathbb{R}^N$ denotes the semi-discrete solution
vector; $\mathbf{M} \in \mathbb{R}^{N \times N}$ is a mass-like matrix; $\mathbf{f} \in \mathbb{R}^N$ is a nonlinear vector function representing  
a semi-discrete flux or an internal force vector; $\mathbf{g}\in\mathbb{R}^N$ is a nonlinear vector function representing a semi-discrete external force vector, source term, or the effect of
some time-dependent boundary conditions; and $\mathbf{u}^0 \in \mathbb{R}^N$ denotes an initial condition for the semi-discrete solution vector. We refer to this problem as
the HDM problem (or simply HDM) and consider its solution using an {\it implicit} time-integration scheme. We refer to the computed solution as the high-dimensional solution (or simply HDM solution). 
We note that this HDM encompasses counterparts associated with nonlinear, semi-discrete, second-order dynamical systems 
characterized by arbitrarily complex dissipative (damping) forces, as such problems can be rewritten in the first-order form~(\ref{eq:hdm}) using a change of variables and some mathematical 
transformations \cite{farhat2014dimensional}.

\subsection{The LSPG method for the dimensionality reduction of a CFD model}
\label{sec:LSPG}

The traditional affine subspace approximation associated with traditional PMOR can be expressed as
\begin{equation}
\begin{aligned}
	\mathbf{u}(t; \mubold) \approx \tilde{\mathbf{u}}(t; \mubold) = \mathbf{u}_\text{ref} + \mathbf{V} \mathbf{q}(t; \mubold)
\end{aligned}
\label{eq:conventional}
\end{equation}
where $\mathbf{u}_{\text{ref}}$ is a reference solution;
and the right ROB $\mathbf{V}$ is constructed as usual --
that is, by adaptively sampling $\mathcal{D} \times (0, T_f]$ using a greedy procedure, solving the HDM problem at the samples 
$\left\{\left(t^i; \mubold^{j}\right)\right\}_{i = 1; \, j = 1}^{i = N_t; \, j = N_{\mu}}$, collecting the 
$N_s = N_t \times N_{\mu}$ computed solution snapshots in the snapshot matrix $\mathbf{S} \in \mathbb{R}^{N\times N_s}$; compressing this matrix using the thin SVD 
$\mathbf{S} = \mathbf{U}_{\mathbf{S}} \mathbf{\Sigma}_{\mathbf{S}} \mathbf{Y}_{\mathbf{S}}^T$, where $\mathbf{U}_{\mathbf{S}} \in \mathbb{R}^{N \times k}$ is the matrix of left singular vectors, 
$\mathbf{\Sigma}_{\mathbf{S}} \in \mathbb{R}^{k \times k}$ is the diagonal matrix of nonzero singular values $\sigma_{\mathbf{S}, i}, i \in \left\{1, \dots, k\right\}$, stored in the order 
$\sigma_{\mathbf{S}, 1} \geq \sigma_{\mathbf{S}, 2} \geq \dots \sigma_{\mathbf{S}, k} > 0$, $\mathbf{Y}_{\mathbf{S}} \in \mathbb{R}^{N_s \times k}$ is the matrix of right singular vectors,
$k \leq \min \left(N, N_s\right)$ denotes the rank of $\mathbf{S}$, and the superscript $T$ designates the transpose; and identifying $\mathbf{V} \in \mathbb{R}^{N \times n}$ as the first $n \ll N$ 
columns of $\mathbf{U}_{\mathbf{S}}$, where the dimension $n$ is typically determined such that 
\begin{equation}
	1 - \frac{\sum\limits_{i = 1}^n \sigma_{\mathbf{S}, i}^2}{\sum\limits_{j = 1}^k \sigma_{\mathbf{S}, j}^2} \leq \varepsilon_{\mathbf{S}}
\label{eq:singular_value_energy_criteria}
\end{equation}
where $\varepsilon_{\mathbf{S}}$ is a user-defined tolerance.

Substituting~\eqref{eq:conventional} in~\eqref{eq:hdm} leads to the nonlinear, semi-discrete, time-dependent residual equation and its associated initial condition
\begin{equation*}
\begin{aligned}
	\mathbf{r}\left(\tilde{\mathbf{u}}(t; \mubold), t; \mubold\right) = \mathbf{r}\left(\mathbf{u}_{\text{ref}} + \mathbf{V q}(t;\mubold), t; \mubold\right) & = \mathbf{M}(\mubold) \mathbf{V} \mathbf{\dot{q}}(t; \mubold) + \mathbf{f}\left(\mathbf{u}_{\text{ref}} + \mathbf{V} \mathbf{q}(t; \mubold); \mubold\right) - \mathbf{g}\left(t; \mubold\right)\\
	\tilde{\mathbf{u}}^0(\mubold) = \mathbf{u}_{\text{ref}} + \mathbf{V} \mathbf{q}(0; \mubold) & \approx \mathbf{u}^0(\mubold)
\end{aligned}
\end{equation*}
The time-discretization of the first of the above equations by a preferred implicit scheme leads at each time-step $m+1$ to a nonlinear system of algebraic equations of the form
\begin{equation} 
	\mathbf{r}^{m+1}(\mubold) = \mathbf{r}\left(\mathbf{u}_{\text{ref}} + \mathbf{V} \mathbf{q}^{m+1}(\mubold), t^{m+1}; \mubold\right)  = \mathbf{0}
	\label{eq:hdmwconv}
\end{equation}
where $\mathbf{r}^{m+1}\in \mathbb{R}^N$ and the superscript $m+1$ (or $m$) designates here and throughout the remainder of this paper a discrete quantity evaluated at time $t^{m+1}$ (or $t^m$). 
The $N$-dimensional system~\eqref{eq:hdmwconv} is overdetermined as it governs $n \ll N$ unknowns -- namely, the generalized coordinates of the subspace approximation~\eqref{eq:conventional} stored
in $\mathbf{q}^{m+1}$. Thus, the unknowns represented by the reduced-order vector $\mathbf{q}^{m+1}$ are constrained by enforcing the orthogonality of the discrete residual $\mathbf{r}^{m+1}$ to a left 
ROB $\mathbf{W} \in \mathbb{R}^{N \times n}$, which leads to the PROM counterpart of the HDM~\eqref{eq:hdm}
\begin{equation}
\begin{aligned}
	\mathbf{W}^T\mathbf{r}^{m+1}(\mubold) = \mathbf{W}^T\mathbf{r}\left(\mathbf{u}_{\text{ref}} + \mathbf{V} \mathbf{q}^{m+1}(\mubold), t^{m+1}; \mubold\right) = \mathbf{0}
\end{aligned}
\label{eq:consresidual}
\end{equation}

In the case of the LSPG PMOR method, the overdetermined system of equations \eqref{eq:hdmwconv} is solved at each time-step for $\mathbf{q}^{m+1}$
using the Gauss-Newton method. In this case, the left ROB is constructed at each Gauss-Newton iteration $\ell+1$ of each time-step $t^{m+1}$ as follows \cite{carlberg2011efficient, carlberg2013gnat}
\begin{equation} 
	\mathbf{W}^{m+1, \ell+1}(\mubold) = \mathbf{J}^{m+1, \ell}(\mubold) \, \mathbf{V}
	\label{eq:WforLSPG}
\end{equation}
where
\begin{equation} 
	\mathbf{J}^{m+1, \ell}(\mubold) = \mathbf{J}\left(\tilde{\mathbf{u}}^{m+1, \ell}(\mubold); \mubold\right) = 
			    \pdv {\mathbf{r}}{\tilde{\mathbf{u}}}\left(\tilde{\mathbf{u}}^{m+1, \ell}(\mubold); \mubold\right) = \pdv {\mathbf{r}}{\tilde{\mathbf{u}}}\left({\mathbf{u}_{\text{ref}} +
			    \mathbf{Vq}}^{m+1, \ell}(\mubold); \mubold\right), \quad \mathbf{J}^{m+1, \ell}(\mubold) \in \mathbb{R}^{N \times N}
	\label{eq:jacobian} 
\end{equation}
Then, solving the discrete, nonlinear PROM problem~\eqref{eq:consresidual} becomes equivalent to solving the optimization problem (see \cite{grimberg2020stability} for a mathematical proof)
\begin{equation*}
\begin{aligned}
	\mathbf{q}^{m + 1}(\mubold) = \arg \min_{\mathbf{x} \in \mathbb{R}^n} \left\| \mathbf{r}\left(\mathbf{u}_{\text{ref}} + \mathbf{V x(\mubold)}, t^{m+1}; \mubold\right)\right\|_2^2
\end{aligned}
\label{eq:gaussnewton}
\end{equation*}

\subsection{The ECSW method for the hyperreduction of an LSPG-based PROM}
\label{sec:ECSW}

For many nonlinear problems of practical interest, the complexity of the solution of problem~\eqref{eq:consresidual} -- and for this matter, that of the processing of any nonlinear PROM -- scales
not only with the small dimension $n$ of the PROM, but also with the large dimension $N$ of the HDM. Hyperreduction \cite{ryckelynck2005priori} is one approach for eliminating this computational 
bottleneck. In this work, we choose ECSW \cite{farhat2014dimensional, farhat2015structure} for hyperreducing any constructed nonlinear PROM, because of its track record established 
in \cite{farhat2014dimensional, farhat2015structure, grimberg2021mesh} for many different large-scale applications; and summarize it below.

Let $\mathcal{E} = \left\{\mathbf{e}_i\right\}, i \in \left\{1, \dots, N_e\right\}$ denote the set of $N_e = \left |\mathcal{E}\right|$ mesh entities defining the discretization of the computational 
domain underlying the semi-discrete HDM~\eqref{eq:hdm}. For example, these entities may be elements in the case of FE modeling, or computational (primal or dual) cells in the case of FV
modeling. In all cases, the discrete, nonlinear PROM~\eqref{eq:consresidual} of dimension $n$ can be rewritten in the case of the LSPG PMOR method as
\begin{eqnarray}
	\mathbf{r}_n^{m+1} (\mubold) &=& \mathbf{W}^{{m+1}^T}(\mubold)\mathbf{r}\left(\mathbf{u}_{\text{ref}} + \mathbf{V} \mathbf{q}^{m+1}(\mubold), t^{m+1}; \mubold\right) \nonumber \\
	&=& \sum_{e_i \in \mathcal{E}} \left(\mathbf{L}_{e_i} \mathbf{W}^{m+1}(\mubold) \right)^T \mathbf{r}_{e_i}\left(\mathbf{L}_{{e_i}^+} \left[\mathbf{u}_{\text{ref}} + \mathbf{V} \mathbf{q}^{m+1}(\mubold)\right], t^{m+1}; \mubold\right)
\label{eq:reducedResidual}
\end{eqnarray}
where:
\begin{itemize}
	\item The notation $\mathbf{\dagger} \left [\bullet \right ]$ is used here and throughout the remainder of this paper to indicate that the quantity $\dagger$ is to be multiplied by the quantity 
		$\bullet$ and not that $\dagger$ is a function of $\bullet$.
	\item $\mathbf{r}_{e_i} \in \mathbb{R}^{d_{e_i}}$ represents the contribution of the individual mesh entity $e_i$ to the discrete residual $\mathbf{r}$ and $d_{e_i}$ denotes the number of dofs 
		attached to $e_i$.  
	\item $\mathbf{L}_{e_i} \in \left\{0, 1\right\}^{d_{e_i} \times N}$ is a Boolean matrix selecting the mesh entity $e_i$ to which $d_{e_i}$ dofs are attached.  
	\item $\mathbf{L}_{{e_i}^+} \in \left\{0, 1\right\}^{d_{{e_i}^+} \times N}$ is a Boolean matrix selecting the union of the mesh entity $e_i$ and its neighbors participating in the same stencil
		of the chosen semi-discretization scheme, to which $d_{{e_i}^+}$ dofs are attached.
	\item The left ROB $\mathbf{W}^{m+1}$ is given by~\eqref{eq:WforLSPG} and~\eqref{eq:jacobian}.
\end{itemize} 

The main idea underlying ECSW is to approximate a projected quantity such as~\eqref{eq:reducedResidual} using a cubature approach whose complexity is independent of $N$, as follows

\begin{eqnarray} 
	\mathbf{r}_n^{m+1} (\mubold) &=& \mathbf{W}^{{m+1}^T}(\mubold)\mathbf{r}\left(\mathbf{u}_{\text{ref}} + \mathbf{V} \mathbf{q}^{m+1}(\mubold), t^{m+1}; \mubold\right)  \nonumber\\
	&\approx& \tilde{\mathbf{r}}_n^{m+1}(\mubold) =  \sum_{e_i \in \widetilde{\mathcal{E}}} \xi_{e_i} \left(\mathbf{L}_{e_i} \mathbf{W}^{m+1}(\mubold) \right)^T \mathbf{r}_{e_i}\left(\mathbf{L}_{{e_i}^+} 
	\left[\mathbf{u}_{\text{ref}} + \mathbf{V} \mathbf{q}^{m+1}(\mubold)\right], t^{m+1}; \mubold\right) 
	\label{eq:hyperreducedResidual}
\end{eqnarray}
where $\widetilde{\mathcal{E}}$ is a subset of $\mathcal{E}$ defining a ``reduced mesh'' -- that is, $n_e = \left |\widetilde{\mathcal{E}} \right| \ll N_e = \left |\mathcal{E}\right |$ -- and 
interpretable as the set of points of the cubature; and $\left \{\xi_{e_1}, \dots, \xi_{e_{n_e}}\right\}$ is the set of {\it positive}, real-valued weights of the cubature approximation.

For implicit time-discretizations, the Jacobian with respect to $\tilde{\mathbf{u}}$ of a discrete, nonlinear, LSPG-based PROM of dimension $n$ and the form given in~\eqref{eq:consresidual} can be written as
\begin{eqnarray*}
	\mathbf{J}_n^{m+1} (\mubold) &=& \mathbf{W}^{{m+1}^T}(\mubold)\mathbf{J}^{m+1}(\mubold) = \mathbf{W}^{{m+1}^T}(\mubold) \frac{\partial \mathbf{r}\left(\mathbf{u}_{\text{ref}} +
	\mathbf{V} \mathbf{q}^{m+1}(\mubold), t^{m+1}; \mubold\right)} {\partial \tilde{\mathbf{u}}}\nonumber \\ 
		&=& \sum_{e_i \in \mathcal{E}} \left(\mathbf{L}_{e_i} \mathbf{W}^{m+1}(\mubold) \right)^T \mathbf{J}_{e_i}\left(\mathbf{L}_{{e_i}^+} \left[\mathbf{u}_{\text{ref}} + \mathbf{V} 
		\mathbf{q}^{m+1}(\mubold) \right]; \mubold\right) 
		\label{eq:reducedJacobian}
\end{eqnarray*}
where $\mathbf{J}_{e_i}\in \mathbb{R}^{d_{e_i}\times d_{e_i}}$ represents the contribution of the individual mesh entity $e_i$ to the Jacobian $\mathbf{J}$ and $d_{e_i}$ denotes as 
before the number of dofs attached to $e_i$. Since $\displaystyle{\mathbf{W}^{{m+1}^T} \partial{\mathbf{r}^{m+1}}/\partial{\tilde{\mathbf{u}}} = 
\partial{\left(\mathbf{W}^{{m+1}^T}\mathbf{r}^{m+1}\right)}/\partial{\tilde{\mathbf{u}}}}$, it follows that $\mathbf{J}_n^{m+1}$ is the Jacobian of the projected 
residual~\eqref{eq:reducedResidual}. Therefore, $\mathbf{J}_n^{m+1}$ can be approximated using the same cubature rule as in~\eqref{eq:hyperreducedResidual}, as follows
\begin{eqnarray} 
	\mathbf{J}_n^{m+1} (\mubold) &=& \mathbf{W}^{{m+1}^T}(\mubold)\mathbf{J}^{m+1}(\mubold) = \mathbf{W}^{{m+1}^T}(\mubold)  
	\frac{\partial \mathbf{r}\left(\mathbf{u}_{\text{ref}} + \mathbf{Vq}^{m+1}(\mubold), t^{m+1}; \mubold\right)}{\partial \tilde{\mathbf{u}}}\nonumber \\
	&\approx& \sum_{e_i \in \widetilde{\mathcal{E}}} \xi_{e_i} \left(\mathbf{L}_{e_i} \mathbf{W}^{m+1}(\mubold) \right)^T \mathbf{J}_{e_i}\left(\mathbf{L}_{{e_i}^+} 
	\left[\mathbf{u}_{\text{ref}} + \mathbf{V} \mathbf{q}^{m+1}(\mubold)\right]; \mubold\right) 
	\label{eq:hyperreducedJacobian}
\end{eqnarray}

Specifically, ECSW computes the reduced mesh $\widetilde{\mathcal{E}}$ and its associated set of weights $\left \{\xi_{e_1}, \dots, \xi_{e_{n_e}}\right\}$ using a machine learning 
approach. Essentially, it trains the approximation~\eqref{eq:hyperreducedResidual} -- from which the counterpart approximation~\eqref{eq:hyperreducedJacobian} can be 
derived --  on a subset of the solution snapshots already collected in the snapshot matrix $\mathbf{S}$, as described below.

Let $\mathbf{C} = [c_{l e}] \in \mathbb{R}^{N_h \times N_e} $ and $\mathbf{d} = [d_l] \in \mathbb{R}^{N_h}$ be the matrix and vector defined as follows
\begin{equation}
\begin{aligned}
	c_{l e} & = \left(\mathbf{L}_e \mathbf{W}^{m+1}\right)^T \mathbf{r}_e\left(\mathbf{L}_{e^+} \left[\mathbf{V V}^T\left( \mathbf{u}^{l} - \mathbf{u}_\text{ref} \right) + 
	\mathbf{u}_\text{ref}\right], (t; \mu)^l\right), 
	& l = 1, \dots, N_h \\ 
	d_l & = \sum_{e \in \mathcal{E}} c_{l e}, & l = 1, \dots, N_h
\end{aligned}
\label{eq:training}
\end{equation}
where $(t; \mu)^l$ is a generic notation for a sampled pair $\left(t^i, \mubold^j\right)$ of a time-step and a parameter point,
and $N_h \le N_s$ denotes the size of the subset of the solution snapshots collected in $\mathbf{S}$ chosen for training the cubature approximation~\eqref{eq:hyperreducedResidual}.
Note that in~\eqref{eq:training} above: 
\begin{itemize}
	\item $\mathbf{V V}^T\left( \mathbf{u}^l - \mathbf{u}_{\text{ref}} \right)$ is the orthogonal projection of the collected solution snapshot $\mathbf{u}^l$ 
		onto the subspace represented by $\mathbf{V}$. This emphasizes that the ECSW training is performed for PROM predictions and not HDM predictions, 
		consistently with~\eqref{eq:hyperreducedResidual}.
	\item $\mathbf{C}$ and $\mathbf{d}$ verify 
		\begin{equation} 
			\mathbf{C} \left[\boldsymbol{\xi} = \mathbf{1}\right] = \mathbf{d} 
			\label{eq:perfect}
		\end{equation}
		where $\boldsymbol{\xi} \in \mathbb{R}^{N_e}$ is the vector of weights of the cubature assuming that $\widetilde{\mathcal E} = \mathcal{E}$ and 
		$\mathbf{1} \in \mathbb{R}^{N_e}$ is the $N_e$-long vector with each entry equal to $1$. This identity is a rewriting of~\eqref{eq:reducedResidual} in matrix form.
\end{itemize}
Then, ECSW can be simply described as relaxing~\eqref{eq:perfect} and defining instead $\widetilde{\mathcal E}$ and $\boldsymbol{\xi}$ as the solution
of the optimization problem
\begin{equation}
	\min_{\boldsymbol{\xi} \in \mathbb{R}^{N_e}_{\geq 0}} \|\boldsymbol{\xi}\|_{\text{ref}} \quad s.t. \quad \|\mathbf{C}\mathbf{\xi} - \mathbf{d}\|_2 \leq \tau\|\mathbf{d}\|_2
	\label{eq:hard}
\end{equation}
where $\|\bullet\|_0$ designates the $\ell_0$-norm of $\bullet$ and $0 < \tau < 1$ is a specified training tolerance. Unfortunately, the above minimization problem is combinatorially hard. For this 
reason, several alternative formulations of~\eqref{eq:hard} were considered in \cite{chapman2017accelerated} and the following nonnegative least squares approach 

\begin{equation}
\begin{aligned}
\boldsymbol{\xi} = \arg \min_{\boldsymbol{\zeta} \in \mathbb{R}_{\geq 0}^{N_e}} \left\| \mathbf{C} \boldsymbol{\zeta} - \mathbf{d} \right\|_2^2
\label{eq:nnls_ecsw}
\end{aligned}
\end{equation}
equipped with the early termination criterion
\begin{equation}
\left \| \mathbf{C} \boldsymbol{\xi} - \mathbf{d} \right\| \leq \tau \left\| \mathbf{d} \right\|_2
\label{eq:nnls_ecsw_early_termination}
\end{equation}
was found to be most practical and computationally efficient. Due to~\eqref{eq:nnls_ecsw_early_termination}, the solution of the optimization problem~\eqref{eq:nnls_ecsw} is a sparse vector
$\boldsymbol{\xi}$ characterized by a relatively small number of nonzero entries corresponding to the mesh entities defining the reduced mesh $\widetilde{\mathcal{E}}$.

After ECSW has computed the reduced mesh $\widetilde{\mathcal E} \subset \mathcal{E}$, the augmented counterpart $\widetilde{\mathcal{E}}^+$ is obtained by simply adding to
$\widetilde{\mathcal{E}}$ the neighbors to its mesh entities that participate in the same stencil of the chosen semi-discretization scheme and have not already been sampled in $\widetilde{\mathcal{E}}$;
and the LSPG-based HPROM associated with an LSPG-based PROM of the form given in~\eqref{eq:consresidual} is computed as
\begin{equation*}
	\label{eq:hprom}
		\mathbf{W}^{{m+1}^T}(\mubold)\tilde{\mathbf{r}}_n^{m+1}(\mubold) = \mathbf{0}
\end{equation*}
where $\tilde{\mathbf{r}}_n^{m+1}$ is given in~\eqref{eq:hyperreducedResidual}.
 
\section{Construction of a nonlinear approximation manifold using a ROB and an ANN}
\label{sec:NONLA}

We propose here a concept of an ANN-augmented PROM -- or a PROM-ANN -- that is grounded in that of an ANN-augmented affine approximation, which itself is inspired from our previous work
on the development of a quadratic approximation manifold~\cite{barnett2022quadratic}. We start from the same snapshot matrix $\mathbf{S} \in \mathbb{R}^{N\times N_s}$ used to construct
a traditional PROM based on the traditional affine subspace approximation. However, we aim at building a right ROB $\mathbf V$ whose dimension $n$ not only verifies $n \ll N$, but is also much smaller
than that of a traditional PROM. In practice, this means that we build $\mathbf{V}$ by using a much looser tolerance $\varepsilon_{\mathbf S}$ in the singular value energy trunction 
criterion~\eqref{eq:singular_value_energy_criteria} than usual. 
To remedy the inaccuracy that can be expected from such a decision, we augment the corresponding affine subspace approximation with an arbitrarily 
nonlinear ANN approximation. Furthermore, we design the ANN approximation such that the size of the corresponding network does not scale with the large dimension $N$ of the parametric HDM of interest, 
but with a much smaller dimension.

Specifically, we construct an arbitrarily nonlinear approximation manifold by approximating the solution $\mathbf{u} \in \mathbb{R}^{N\times n}$ as follows
\begin{equation}
	\mathbf{u}(t; \mubold) \approx \tilde{\mathbf{u}}(t; \mubold) = \mathbf{u}_\text{ref} + \mathbf{V} \mathbf{q}(t; \mubold) + \overline{\mathbf{V}}\mathcal{N}(\mathbf{q}(t; \mubold))
	\label{eq:ROB-ANN}
\end{equation}
where:
\begin{itemize}
	\item $\mathbf{V} \in \mathbb{R}^{N \times n}$ is constructed as usual, using the first $n \ll N$ columns of $\mathbf{U}_{\mathbf{S}}$ -- the matrix of left singular vectors of the snapshot
		matrix $\mathbf{S}$ (see Section \ref{sec:LSPG}) which spans the range of $\mathbf{S}$ -- with $n$ also much smaller than for a traditional PROM.
	\item $\overline{\mathbf{V}} \in \mathbb{R}^{N \times \bar n}$ is constructed using a subset of the next $\bar n \ll N$ columns of  $\mathbf{U}_{\mathbf{S}}$, and therefore $\bar n \le k - n$,
		where $k \leq \min \left(N, N_s\right)$ is the rank of $\mathbf{S}$ (see Section \ref{sec:LSPG}).
	\item $\mathbf{V}$ and $\overline{\mathbf{V}}$ satisfy the orthogonality properties
		\begin{equation}
			\label{eq:ORTH}
			\mathbf{V}^T\mathbf{V} = \mathbf{I}_n, \qquad \overline{\mathbf{V}}^T\overline{\mathbf{V}} = \mathbf{I}_{\bar n}, 
			\qquad \mathbf{V}^T\overline{\mathbf{V}} = \mathbf{0}, \quad \overline{\mathbf{V}}^T\mathbf{V} = \mathbf{0}
		\end{equation}
		where $\mathbf{I}_n$ and $\mathbf{I}_{\bar n}$ denote the identity matrices of dimension $n$ and $\bar n$, respectively.
	\item $\mathcal{N}: \mathbb{R}^n \to \mathbb{R}^{\bar n}$ is a map constructed by an ANN whose size scales with $\max(n, \bar n) \ll N$.
\end{itemize}

In practice, we emphasize in~\eqref{eq:ROB-ANN} the contribution of the ANN and therefore choose $n$ and $\bar n$ such that $n \ll \bar n$. This is because only the contribution of 
$\overline{\mathbf{V}}\mathcal{N}$ to the overall approximation~\eqref{eq:ROB-ANN} can effectively mitigate the Kolmogorov barrier to reducibility. 
That of the affine subspace approximation is to provide ``completeness'' to the overall approximation~\eqref{eq:ROB-ANN}.

To characterize further the ANN representation $\mathcal N$, we consider the projection of all solution snapshots computed and collected in the snapshot matrix $\mathbf{S}$ (or a subset of them) onto 
the subspaces represented by $\mathbf V$ and $\overline{\mathbf V}$. They can be written as
\begin{equation}
	\label{eq:PROJ}
	\mathbf{q}^l = \mathbf{V}^T\left(\mathbf{u}^l - \mathbf{u}_\text{ref}\right), \quad  \bar{\mathbf{q}}^l = \overline{\mathbf{V}}^T\left(\mathbf{u}^l - \mathbf{u}_\text{ref}\right), 
	\quad l = 1, \dots, N_s
\end{equation}
Next, we apply~\eqref{eq:ROB-ANN} to approximate ultimately perfectly all computed solution snapshots, which yields
\begin{equation}
	\label{eq:APSN1}
	\mathbf{u}^l(\mubold) - \mathbf{u}_\text{ref} = \mathbf{V} \mathbf{q}^l + \overline{\mathbf{V}}\mathcal{N}\left(\mathbf{q}^l\right)
\end{equation}
We premultiply~\eqref{eq:APSN1} by $\overline{\mathbf{V}}^T$, which in view of the orthogonality properties~\eqref{eq:ORTH} leads to
\begin{equation*}
	\label{eq:APSN2}
	\overline{\mathbf V}^T\left(\mathbf{u}^l - \mathbf{u}_\text{ref}\right)= \mathcal{N}\left(\mathbf{q}^l\right)
\end{equation*}
and in view of~\eqref{eq:PROJ} gives
\begin{equation} 
	\label{eq:APSN3} 
	\bar{\mathbf{q}}^l =  \mathcal{N}\left(\mathbf{q}^l\right), \quad l = 1, \dots, N_s
\end{equation}

Hence, our ANN takes as input the data set $\left\{\mathbf{q}^l \right\}_{l = 1}^{N_s}$ (or a subset of it); and produces the output data set $\left\{\bar{\mathbf{q}}^l \right\}_{l = 1}^{N_s}$ (or a
subset of it). We determine its vector-valued hyperparameter $\alphabold$ by minimizing the loss function implied by~\eqref{eq:NN_opt_problem} -- that is,
\begin{equation}
	\alphabold = \arg \min_{\alphabold^{\prime}} \displaystyle{\frac{1}{N_\text{train}} \sum_l^{N_\text{train}} \left(\bar{\mathbf{q}}^l - \mathcal{N}\left(\mathbf{q}^l, \alphabold^\prime\right) \right)^2}
	\label{eq:NN_opt_problem}
\end{equation}
where $N_\text{train} < N_s$ denotes the number of solution snapshots allocated to training.

\section{Adaptation of LSPG and ECSW to the construction of an HPROM-ANN}
\label{sec:ADAP}

\subsection{Adaptation of LSPG}

Given a solution approximation such as~\eqref{eq:conventional} or~\eqref{eq:ROB-ANN}, LSPG is fully specified once its left ROB $\mathbf{W}$ is specified. Hence, we adapt here LSPG to the
proposed concept of an ANN-augmented PROM by fully specifying the left ROB $\mathbf{W}$ associated with the approximation~\eqref{eq:ROB-ANN}.

To this end, we first substitute in~\eqref{eq:hdmwconv} the ANN-augmented affine approximation~\eqref{eq:ROB-ANN} for the traditional affine subspace approximation, which yields
\begin{equation} 
	\label{eq:NPR} 
	\mathbf{r}\left(\underbrace{\mathbf{u}_{\text{ref}} + \mathbf{Vq}^{m+1}(\mubold) + \overline{\mathbf{V}}\mathcal{N}\left({\mathbf{q}^{m+1}(\mubold)}\right)}_
	{{\tilde{\mathbf{u}}}\left(\mathbf{q}^{m+1}(\mubold); \mubold\right)}, t^{m+1}; \mubold\right) = \mathbf{0}
\end{equation}

Next, we apply the Gauss-Newton method to solve the above overdetermined system of $N$ equations with $n\ll N$ unknowns $\left(\mathbf{q}^{m+1} \in \mathbb R^n\right)$, which can be summarized as follows: 
\begin{itemize}
	\item Build the sequence of iterations 
		\begin{equation} 
			\label{eq:INC} 
			\mathbf{q}^{m+1, \ell + 1}(\mubold) = \mathbf{q}^{m+1, \ell}(\mubold) + \Delta \mathbf{q}^{m+1, \ell+1}(\mubold) 
		\end{equation} 
		where $\mathbf{q}^{m+1, \ell}$ denotes the $\ell$-th iterate generalized coordinate solution. 
	\item Substitute~\eqref{eq:INC} into~\eqref{eq:NPR} and perform the first-order Taylor series approximation of the result around $\mathbf{q}^{m+1, \ell}$ -- which can be written as
		\begin{align} 
			\mathbf{r} & \left(\tilde{\mathbf{u}}^{m+1}\left(\mathbf{q}^{m+1, \ell}(\mubold) + \Delta \mathbf{q}^{m+1, \ell+1}(\mubold); \mubold\right), t^{m+1}; \mubold \right) \approx 
			\mathbf{r} \left(\tilde{\mathbf{u}}\left(\mathbf{q}^{m+1, \ell}(\mubold); \mubold\right), t^{m+1}; \mubold\right) \nonumber\\ 
			&\, + \mathbf{J}^{m+1, \ell}(\mubold)\pdv{\tilde{\mathbf{u}}}{\mathbf{q}}\left(\mathbf{q}^{m+1, \ell}
			(\mubold)\right)\Delta \mathbf{q}^{m+1, \ell+1}(\mubold) = \mathbf{0} 
			\label{eq:TS0} 
		\end{align} 
		where 
		\begin{equation*}
			\label{eq:TS1}
			\mathbf{J}^{m+1, \ell}(\mubold) = \mathbf{J}\left(\tilde{\mathbf{u}}\left(\mathbf{q}^{m+1, \ell}(\mubold); \mubold\right)\right)
		\end{equation*}
		and
		\begin{equation} 
			\displaystyle{\frac{\partial \tilde{\mathbf{u}}}{\partial \mathbf{q}}\left(\mathbf{q}^{m+1, \ell}(\mubold)\right) = 
			\frac{\partial \left(\mathbf{Vq}^{m+1, \ell}(\mubold) + \overline{\mathbf{V}}{\mathcal N}\left(\mathbf{q}^{{m+1, \ell}}(\mubold)\right)\right)}{\partial \mathbf{q}}} 
			= \mathbf{V} + \overline{\mathbf{V}} \displaystyle{\frac{\partial{\mathcal N}}{\partial \mathbf{q}}\left(\mathbf{q}^{m+1, \ell}(\mubold)\right)}
			\label{eq:TS2} 
		\end{equation}
	\item Rearrange equation~\eqref{eq:TS0} as
		\begin{equation} 
			\mathbf{W}^{m+1, \ell+1}(\mubold)\,\Delta \mathbf{q}^{m+1, \ell+1}  = 
			- \mathbf{r} \left(\tilde{\mathbf{u}}\left(\mathbf{q}^{m+1, \ell}(\mubold)\right), t^{m+1}; \mubold\right) 
			\label{eq:TS3} 
		\end{equation} 
		where
		\begin{equation*}
			\mathbf{W}^{m+1, \ell+1}(\mubold) = \mathbf{J}^{m+1, \ell}(\mubold) \pdv{\tilde{\mathbf{u}}}{\mathbf{q}} \left(\mathbf{q}^{m+1, \ell}(\mubold)\right)
			\label{eq:TS4}
		\end{equation*}
		which in view of~\eqref{eq:TS2} can also be written as
		\begin{equation} 
			\mathbf{W}^{m+1, \ell+1}(\mubold) = \mathbf{J}^{m+1, \ell}(\mubold)
			\left(\mathbf{V} + \overline{\mathbf{V}} \displaystyle{\frac{\partial{\mathcal N}}{\partial \mathbf{q}}\left(\mathbf{q}^{m+1, \ell}(\mubold)\right)}\right)
			\label{eq:TS6}
		\end{equation}
\end{itemize}

The expression of the left ROB given in~\eqref{eq:TS6} specifies the left ROB associated with the ANN-augmented affine approximation~\eqref{eq:ROB-ANN} and therefore
completes the description of the LSPG method for the proposed concept of an ANN-augmented PROM. 

From~\eqref{eq:WforLSPG} and~\eqref{eq:TS6}, it follows that in the case of the ANN-augmented affine approximation, the left ROB of LSPG is simply obtained by adding 
$$\mathbf{W}^{m+1, \ell+1}_{\mathcal N}(\mubold) = \mathbf{J}^{m+1, \ell}(\mubold) \overline{\mathbf{V}} \displaystyle{\frac{\partial{\mathcal N}} 
{\partial \mathbf{q}}\left(\mathbf{q}^{m+1, \ell}(\mubold)\right)}$$
to the left ROB 
\begin{equation*}
	\mathbf{W}^{m+1, \ell+1}_\text{tra}(\mubold) = \mathbf{J}^{m+1, \ell}(\mubold)\mathbf{V}
\end{equation*}
computed in the case of the traditional affine subspace approximation. This incurs at each Gauss-Newton iteration primarily two additional 
computational costs: that associated with the evaluation of the gradient $\partial \mathcal{N}/\partial {\mathbf{q}}$ ($C_1$); and that associated with the left multiplication of this gradient by 
the right ROB $\overline{\mathbf{V}}$ ($C_2$). It also entails an additional storage requirement.

Regarding the additional cost $C_1$, we reemphasize that the map $\mathcal{N}: \mathbb{R}^n \to \mathbb{R}^{\bar n}$ is constructed by an ANN whose size scales with $\max(n, \bar n) \ll N$:
hence, the computational cost incurred by the online evaluation of the gradient of this map is negligible compared to those costs associated with building the left ROB 
$\mathbf{W}^{m+1, \ell+1}_\text{tra}$, which scale with $N^{\beta}$, $\beta > 1$. Furthermore, because by design, the proposed ANN has more outputs than inputs ($\bar n \gg n$), we mitigate 
this cost by using the {\it forward} rather than backward differentiation mode adopted during training -- that is, by evaluating this gradient 
with respect to the inputs instead of the outputs.

Regarding the additional cost $C_2$, it is associated with $\mathcal{O} (Nn\bar n)$ additional operations. Hence, it is negligible compared to the cost of building 
$\mathbf{W}^{m+1, \ell+1}_\text{tra}$, which, in the absence of hyperreduction, is associated with $\mathcal{O} (N^3n)$ operations. 

As for the additional storage requirement, it is associated with storing the matrix $\left(\overline{\mathbf{V}} \partial \mathcal{N}/\partial \mathbf{q}\right) \in \mathbb{R}^{N \times n}$. 
Hence, it is negligible compared to the storage requirements associated with the construction of the left ROB $\mathbf{W}^{m+1, \ell+1}_\text{tra}$, because $n \ll N$.

Finally, we note that the standard Gauss-Newton method solves the overdetermined problem~\eqref{eq:TS3} by transforming it into the square problem
\begin{equation} 
	\left(\mathbf{W}^{{m+1, \ell+1}^T}(\mubold)\,\mathbf{W}^{m+1. \ell+1}(\mubold)\right)\,\Delta \mathbf{q}^{m+1, \ell+1}  
	= - \mathbf{W}^{{m+1, \ell+1}^T}\mathbf{r} \left(\tilde{\mathbf{u}}\left(\mathbf{q}^{m+1, \ell}(\mubold)\right), t^{m+1}; \mubold\right) 
	\label{eq:TS5} 
\end{equation}
and solving this problem instead. This requires however guaranteeing that problem~\eqref{eq:TS5} above is nonsingular -- that is, ensuring that the matrix $\mathbf{W}^{m+1, \ell+1}$ has full 
column rank. From~\eqref{eq:TS6}, it follows that this requirement imposes a counterpart requirement on the ANN designed for computing the map $\mathcal N$ whose satisfaction may not be an easy task. 
For this reason, problem~\eqref{eq:TS3} is solved instead in this work using the truncated SVD method, where truncation amounts in this case to regularization 
(for example, see \cite{barnett2022quadratic} for more details).

\subsection{Adaptation of ECSW}

Next, we adapt ECSW to the concept of an ANN-augmented PROM. For this purpose, we first point the attention of the reader to ~\eqref{eq:reducedResidual} and the first bullet below that equation, 
which emphasize that for a given PROM characterized by a ROB $\mathbf V$, ECSW trains the cubature approximation for PROM predictions defined by generalized coordinates and not for HDM counterparts. 
This means that in principle, one should exercise the constructed PROM to compute and collect generalized coordinates solution snapshots, which unfortunately defeats the purpose of hyperreduction. 
In the case of the traditional affine subspace approximation, this issue can be elegantly resolved by simply projecting the high-dimensional solution snapshots collected in the matrix $\mathbf S$, 
or a subset 
of them, onto the right ROB $\mathbf{V}$. This is highlighted in \eqref{eq:training}, which describes the construction of the matrix $\mathbf{C}$ and vector $\mathbf{d}$ defining the least squares 
problem~\eqref{eq:nnls_ecsw} and its early termination criterion~\eqref{eq:nnls_ecsw_early_termination}.

In the case of the proposed mixed ANN-augmented affine approximation~\eqref{eq:ROB-ANN} however, the identification of the vector of generalized coordinates $\mathbf{q}^l$ associated with a collected 
high-dimensional solution snapshot $\mathbf{u}^l$ can no longer be obtained via projection onto either of the ROBs $\mathbf{V}$ or $\overline{\mathbf{V}}$. Instead, it requires the solution of a 
nonlinear problem of the form
\begin{equation*} 
	\boldsymbol{\delta}^l(\mathbf{q}^l) =  \overline{\mathbf{V}} {\mathcal N}\left(\mathbf{q}^l\right) + \mathbf{Vq}^l + \mathbf{u}_{\text{ref}} - \mathbf{u}^l = \mathbf{0} 
\end{equation*}
using, for example, yet another Gauss-Newton procedure that can be summarized as follows
\begin{align*}
	\mathbf{q}^{l,0} & = \mathbf{V}^T \mathbf{u}^l \\ 
	\mathbf{q}^{l, \ell+1} & = \mathbf{q}^{l, \ell} - \left(\pdv{\boldsymbol{\delta}^l}{\mathbf{q}}\left(\mathbf{q}^{l, \ell}\right)\right)^+
	\boldsymbol{\delta}^l\left(\mathbf{q}^{l, \ell}\right)
\end{align*}
where the superscript $\ell$ designates here too the $\ell$-th Gauss-Newton iteration and the superscript $+$ designates the Moore-Penrose inverse. 
It follows that for the same number $N_h$ of training solution snapshots, the computational cost associated with the setup of ECSW 
in the case of our proposed ANN-augmented affine approximation is higher than in the case of the traditional affine subspace approximation.
Nevertheless, we can justify this potential cost increase and anticipate a significant return on our investment using the following observation
(also, see~\cite{barnett2022quadratic}).

Given a nonlinear, convection-dominated flow problem, we can expect the ANN-augmented affine approximation to deliver a similar level of accuracy as a traditional PROM, using however a 
much smaller dimension $n$. Considering this and the fact that the number of cubature points required for approximating a $d$-dimensional integral function with $p$ cubature points along each 
dimension grows as $p^d$, we can reasonably expect ECSW to deliver in the case of the ANN-augmented affine approximation that we propose a much smaller reduced mesh than in that of the traditional 
affine subspace approximation. Hence, we can expect ECSW to deliver an even better computational efficiency when it is applied to hyperreduce a PROM-ANN; 
and expect an ECSW-based HPROM-ANN to deliver an even 
better computational efficiency than a traditional HPROM. The numerical results we report in Section \ref{sec:APP} for a popular, parametric benchmark problem support this conclusion.

\section{Application and performance assessment}
\label{sec:APP}

Here, we illustrate the concept of an ANN-augmented PROM proposed in this paper using a parametric, 2D, inviscid Burgers' problem defined in a bounded computational domain and
a time-interval $ (0, T_f]$. We also contrast the performance of this concept for this problem, after hyperreduction, with that of a traditional HPROM, for a fixed level of accuracy. 

The inviscid Burgers' equation is a popular benchmark for PMOR whose semi-discretization can be written in the form given in~\eqref{eq:hdm}. It is strictly convection-dominated and well known for 
exhibiting the Kolmogorov $n$-width issue. Depending on the specified initial condition, boundary condition, or source term, its solution may develop propagating shocks . 

In all cases, we choose $\mathbf{u}_\text{ref} = \mathbf{0}$. We assess the accuracy of each computed HPROM solution relative to the HDM solution, using the following measure of the relative error
\begin{equation*}
	\mathbb{RE} = \displaystyle{\frac{\sum\limits_{m=0}^{N_\tau} \left|\left|\mathbf{u}^{m} - \tilde{\mathbf{u}}^m\right|\right|_2}{\sum\limits_{m=0}^{N_\tau} \left|\left|\mathbf{u}^m\right|\right|_2}}
	\label{eq:relative_error}
\end{equation*}
where $N_\tau$ denotes the total number of equally spaced time-steps sampled in $(0, T_f]$ for the purpose of computing $\mathbb{RE}$; and $\tilde{\mathbf {u}}^m$ is the corresponding HPROM or 
HPROM-ANN solution at time $m \Delta t$.

We perform all computations that we report on in this section in double precision arithmetic, on a single core of a single node of a Linux cluster, where each node is configured with two Intel Xeon 
Gold 5118 processors clocked at 2.3 GHz and 192 GB of memory. Hence, all solution times we report are wall clock times.

\subsection{Parameterized two-dimensional inviscid Burgers' problem and discretization}

Specifically, we consider the following parametric, 2D, inviscid Burgers' initial boundary value problem (IBVP) defined in a 2D parameter domain $\mathcal D \subset \mathbb{R}^{2}$ 
($N_{\mathcal D} = 2$)
\begin{align}
\label{eq:burgers} 
	\begin{split} 
		\pdv{{u_x}}{t} + \frac{1}{2}\left(\pdv{{u_x}^2}{x} + \pdv{{u_x u_y}}{y} \right) & = 0.02 \exp(\mu_2 x) \\ 
		\pdv{{u_y}}{t} + \frac{1}{2}\left(\pdv{{u_y}^2}{y} + \pdv{{u_x u_y}}{x} \right) & = 0 \\ 
		{u_x}(0, y, t; {\mubold}) & = \mu_1 \\ 
		u_x(x, y, 0) = u_y(x, y, 0) & = 1 
	\end{split} 
\end{align} 
where $x \in [0, 100]$, $y \in [0, 100]$, $t \in (0, T_f]$, $T_f = 25$, and $\mubold = \left(\mu_1, \mu_2\right) \in [4.25, 5.50] \times [0.015, 0.03]$.

We discretize the 2D computational domain $[0, 100] \times [0, 100]$ in $250 \times 250$ equally sized elements, which leads to $N_e = 62\,500$ and $N = 125\,000$. 
We semi-discretize the parametric IBVP on this mesh using Godunov's method. The resulting semi-discrete HDM can be written in the form given in~\eqref{eq:hdm}, where
\begin{equation*}
	\mathbf{u} = \mqty(u_x \\ u_y)
\end{equation*}

We discretize the resulting ordinary differential equation using the trapezoidal rule and the fixed time-step size $\Delta t = 0.05$, and therefore discretize the time-interval $(0, 25]$ in
500 equally space time-steps. For simplicity, we set $N_t = N_\tau = 500$. 

On a single core of the Linux cluster described above, the time-to-solution of a single instance of the parametric HDM is $718$ seconds.

\begin{figure}[!h]
	\centering
	\includegraphics[width=0.85\textwidth]{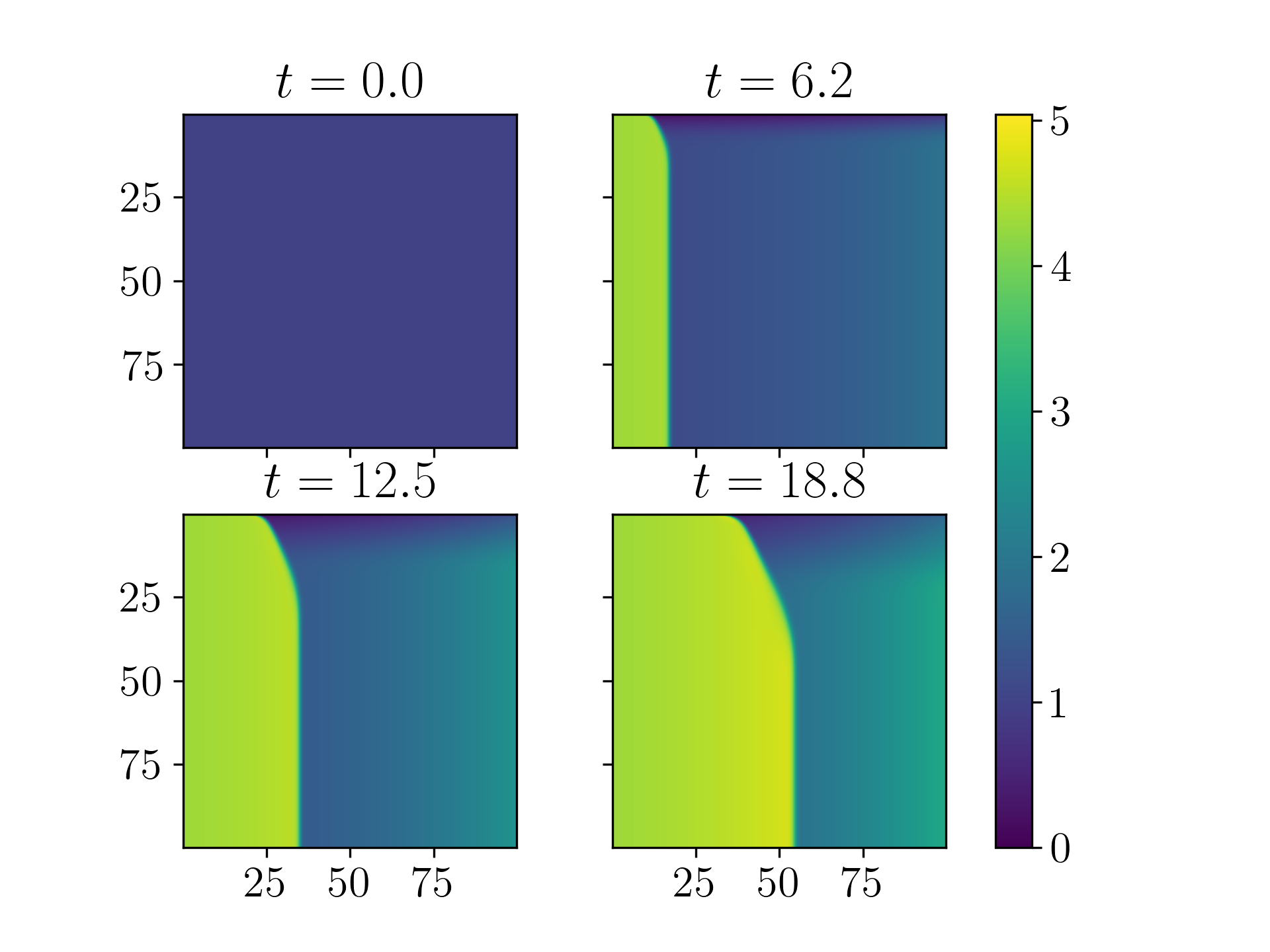}
	\caption{Snapshots of the component $u_x$ of the HDM solution for $\mubold = (4.3, 0.021)$.}
	\label{fig:hdm_example}
\end{figure}

\begin{figure}[!h]
	\centering
	\includegraphics[width=0.85\textwidth]{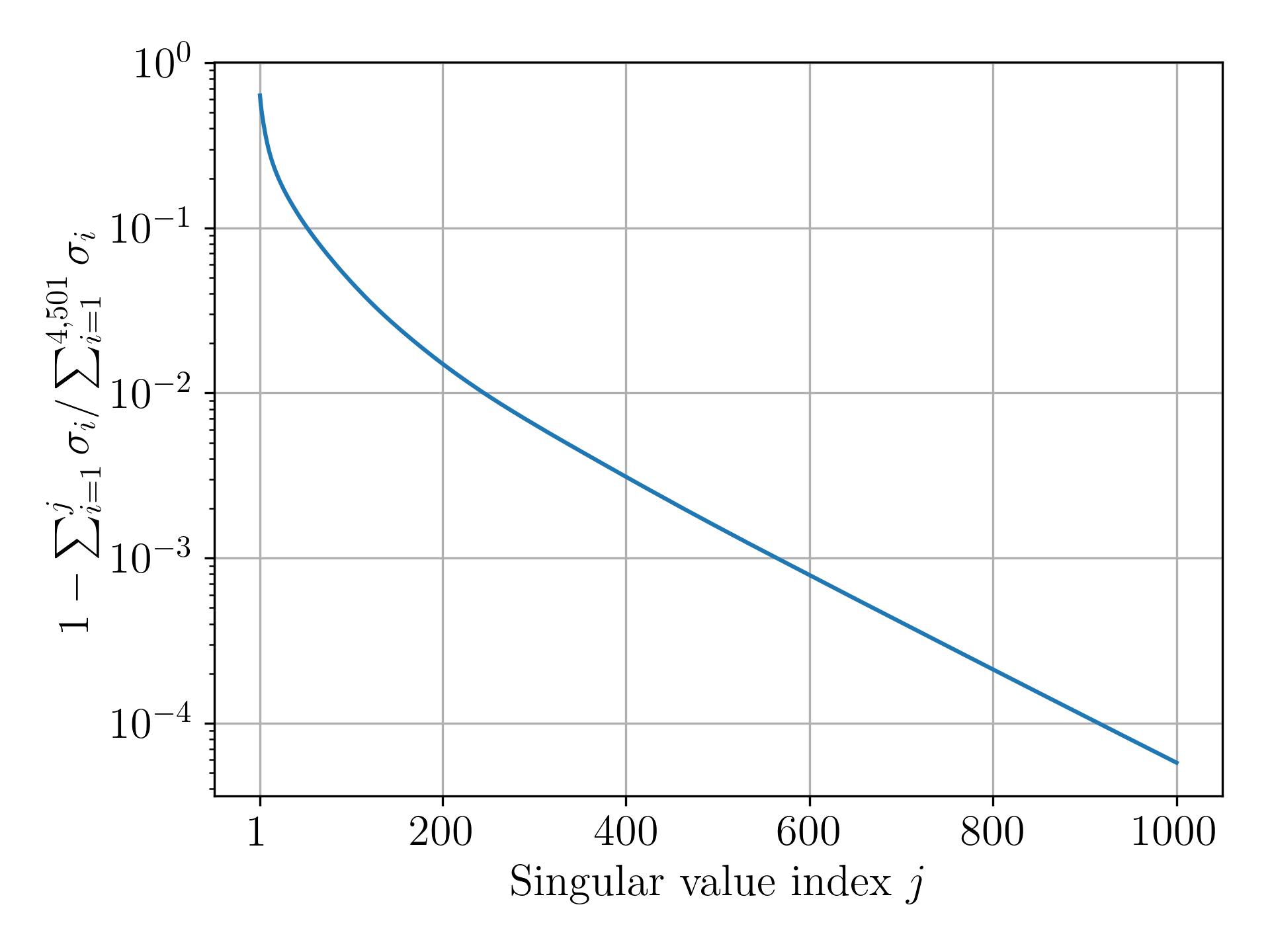}
	\caption{Decay of the singular value energy fraction of the largest 1\,000 singular values of the snapshot matrix.} 
	\label{fig:sing_vals}
\end{figure}

Fig. \ref{fig:hdm_example} shows that the solution of the parametric IBVP~\eqref{eq:burgers} generally produces a shock that propagates from the left to the right of the computational domain. 
This convecting structure leads to the Kolmogorov barrier, which manifests itself in Fig. \ref{fig:sing_vals} through the slow decay of the singular values of the snapshot matrix $\mathbf S$.


\newpage
\subsection{Training the ROB}
\label{sec:TR}

For the purpose of training in both cases of a PROM and a PROM-ANN the ROB $\mathbf V$, we uniformly sample the parameter domain $\mathcal{D} = [4.25, 5.50] \times [0.015, 0.03]$ by a 
$3 \times 3$ grid -- and therefore 9 parameter points -- using the constant increments $\Delta \mu_1 = 0.625$ and $\Delta \mu_2 = 0.0075$. This sampling leads to $N_s = 4\,501$ solution snapshots, 
including the initial condition which is nonparametric and therefore shared by all sampled parameter points (see~\eqref{eq:burgers}). We train ECSW using: only one of the sampled parameter 
points -- namely, the point (4.25, 0.0225); every 10-th solution snapshot in time; and therefore a subset of $N_h = 50 (1 \times 50)$ of the collected $4\,501$ solution snapshots.

\subsection{Parametric nonlinear LSPG- and ECSW-based HPROM}

\begin{figure}[!h]
	\centering
	\includegraphics[width=0.85\textwidth]{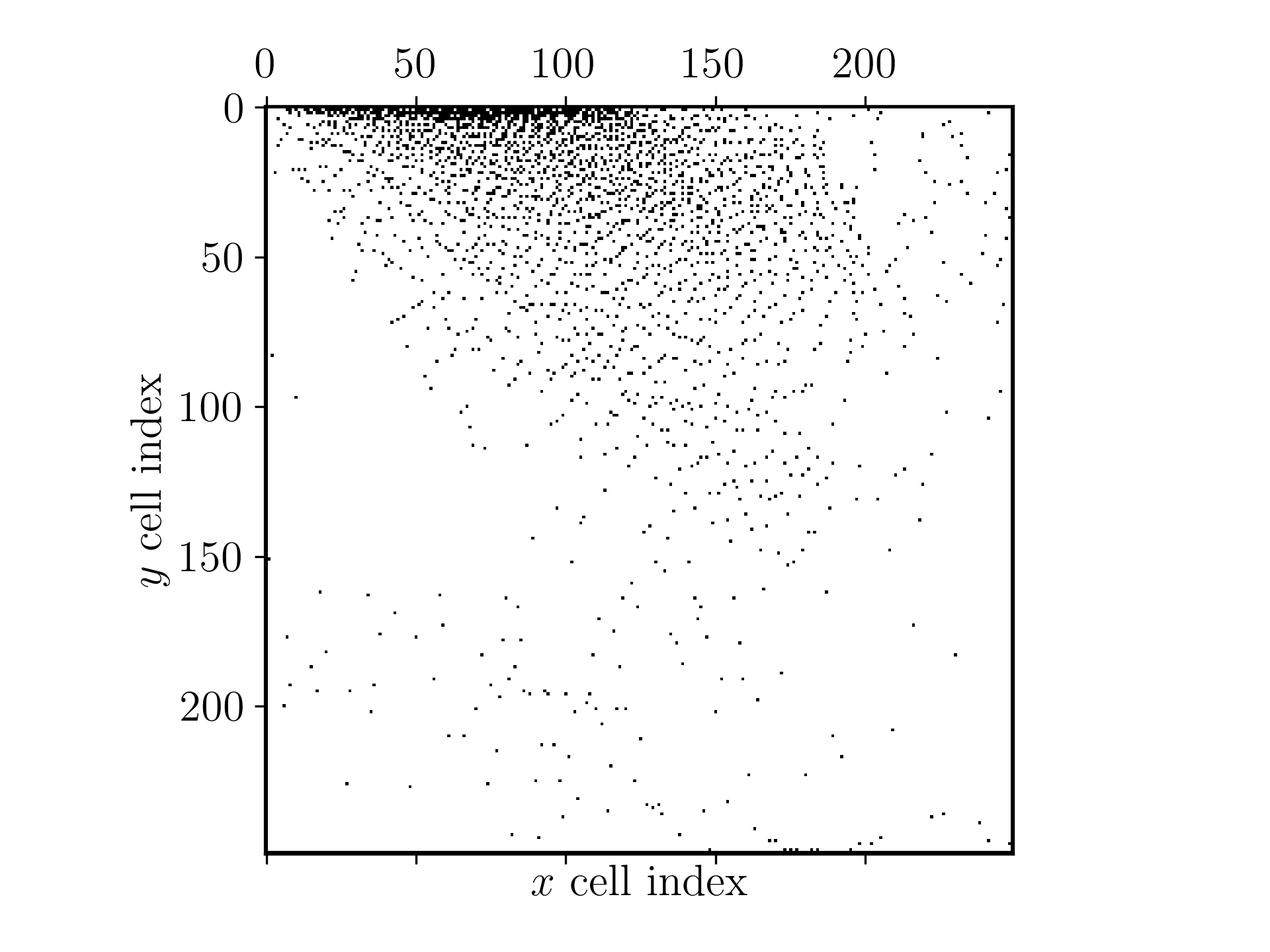}
	\caption{ECSW-generated reduced mesh for the traditional LSPG-based PROM.}
	\label{fig:prom_red_mesh}
\end{figure}

First, we set $\varepsilon_{\mathbf S}$~\eqref{eq:singular_value_energy_criteria} to $5\%$ and construct a right ROB $\mathbf{V}$ of dimension $n = 95$ using the training overviewed 
in Section \eqref{sec:TR}. Using this ROB and the corresponding left ROB $\mathbf{W}$, ECSW delivers 
a reduced mesh with $n_e = 5\,689$ elements ($\ll N_e = 62\,500$ elements) 
(see Table \ref{table:quantitative_performance}). 
This reduced mesh is graphically depicted in Fig. \ref{fig:prom_red_mesh}.

\begin{figure}[!h] 
	\centering 
	\includegraphics[width=0.85\textwidth]{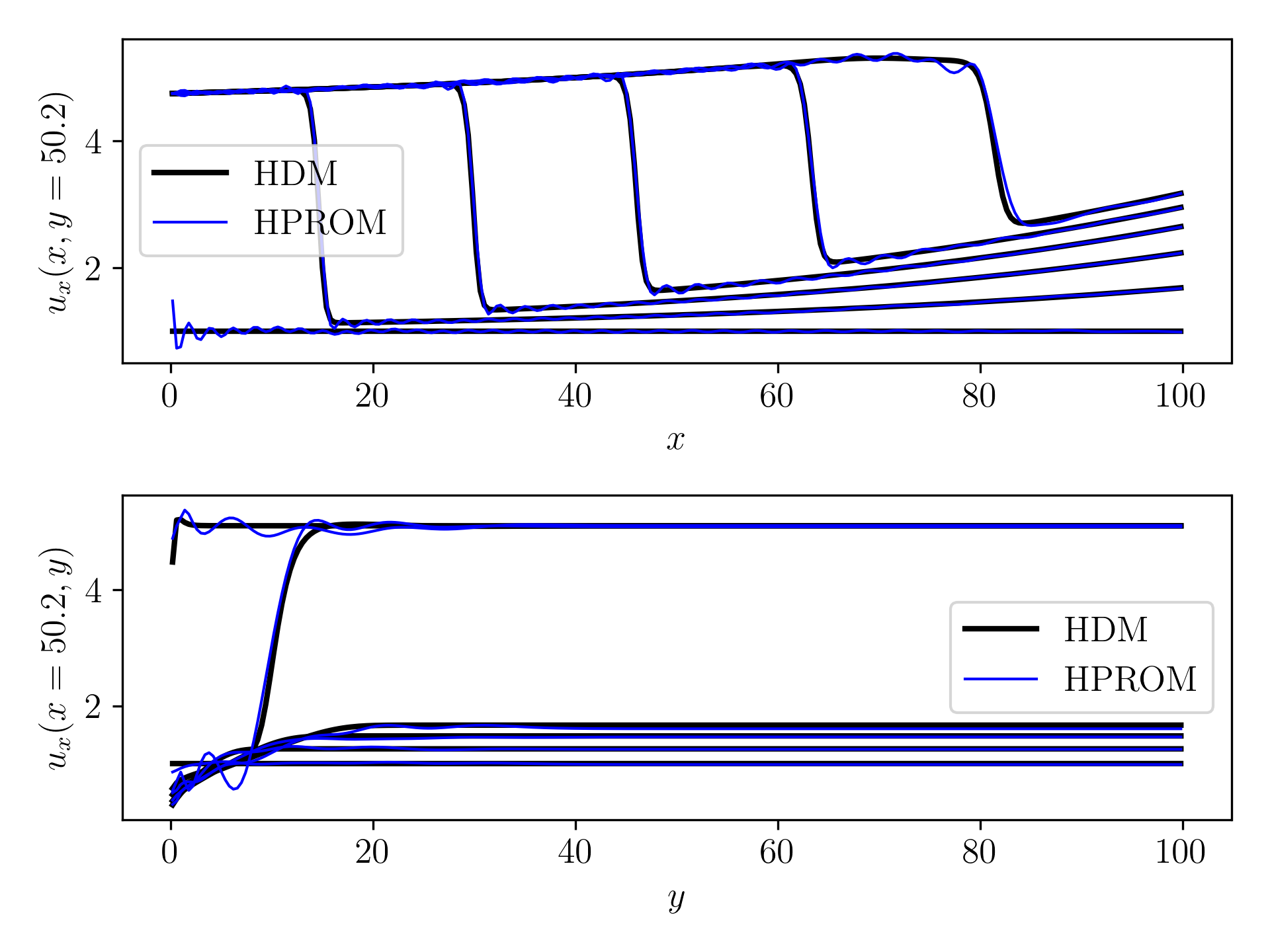} 
	\caption{Numerical predictions performed for $\mubold = (4.75, 0.02)$ using the LSPG- and ECSW-based HPROM ($t = 0, 5, 10, 15, 20,$ and $25$): $y = 50.2$ (top); and $x = 50.2$ (bottom).}
	\label{fig:hprom_plot}
\end{figure}

Next, we query the unsampled parameter point $\mubold = (4.75, 0.02)$ and apply the constructed LSPG- and ECSW-based HPROM to predict the solution of the IBVP~\eqref{eq:burgers}.
We report in Fig. \ref{fig:hprom_plot} the results obtained in two mid-plane slices of the computational domain. These results are in good agreement with their HDM counterparts,
which is confirmed by the relative errors of the order of $1\%$ reported in Table \ref{table:quantitative_performance}.

\newpage
\subsection{Parametric nonlinear LSPG- and ECSW-based HPROM-ANN}

Here, we first set $n = 10$ ($\ll 95$), $\bar{n} = 140$, and construct two right ROBs $\mathbf{V}$ and $\overline{\mathbf V}$ using the training briefly described in Section \eqref{sec:TR}.
Next, we use the python software package \texttt{pytorch} to construct for the map $\mathcal N$ the ANN graphically depicted below
$$ (n, 32) \underset{\text{ELU}}{\rightarrow} (32, 64) \underset{\text{ELU}}{\rightarrow}(64, 128) \underset{\text{ELU}}{\rightarrow} (128, 256) \underset{\text{ELU}}{\rightarrow} (256, 256) \underset{\text{ELU}}{\rightarrow} (256, \overline{n}) $$
This ANN has 6 linear layers of dimensions (input, output) and  exponential linear unit (ELU) activation functions. We split all $N_s = 4\,501$ solution snapshots between training and testing. 
Specifically, we randomly shuffle the set of generalized coordinates associated with these snapshots and split the data 90\%-10\% for training-testing. We also use \texttt{pytorch} to compute the
gradient $\partial \mathcal{N}/\partial {\mathbf{q}}$ (forward mode automatic differentiation).

\begin{figure}[!h]
	\centering
	\includegraphics[width=0.85\textwidth]{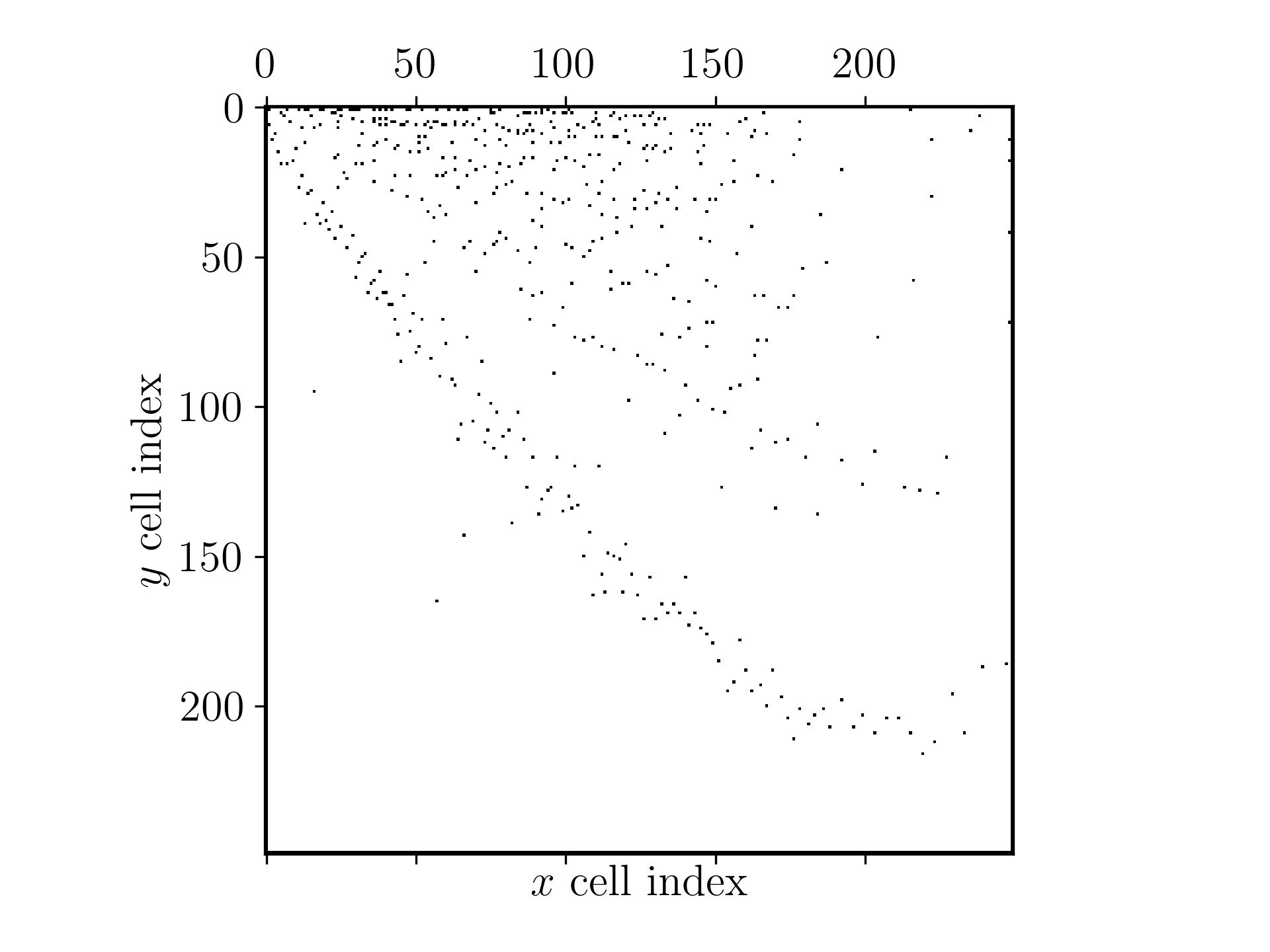}
	\caption{ECSW-generated reduced mesh for the LSPG-based PROM-ANN.}
	\label{fig:prom-nn_red_mesh}
\end{figure}

Next, we use the constructed right ROBs and ANN to build a PROM-ANN and apply ECSW to hyperreduce it, and hence transform it into an HPROM-ANN. In this case, ECSW delivers a reduced mesh 
with $n_e = 1\,279$ elements ($\ll N_e = 62\,500$ elements) (see Table \ref{table:quantitative_performance}). This reduced mesh is shown in Fig. \ref{fig:prom-nn_red_mesh}.

\begin{figure}[!h]
	\centering
	\includegraphics[width=0.85\textwidth]{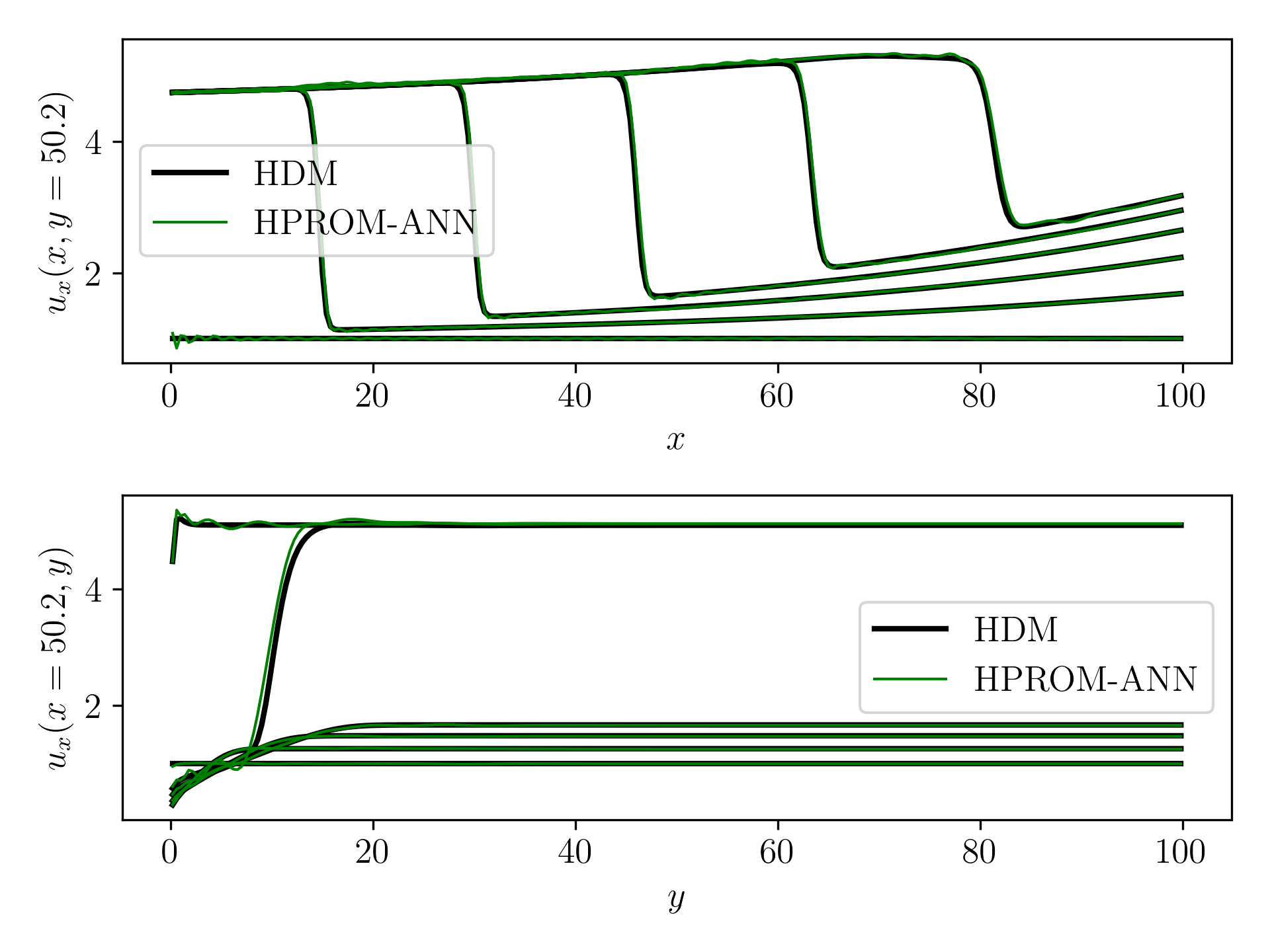}
	\caption{Numerical predictions performed for $\mubold = (4.75, 0.02)$ using the LSPG- and ECSW-based HPROM-ANN ($t = 0, 5, 10, 15, 20,$ and $25$): $y = 50.2$ (top); and $x = 50.2$ (bottom).}
	\label{fig:hprom-nn_plot}
\end{figure}

Then, we query the same unsampled parameter point $\mubold = (4.75, 0.02)$ and apply the constructed LSPG- and ECSW-based HPROM-ANN to predict the solution of the IBVP~\eqref{eq:burgers}.
We report in Fig. \ref{fig:hprom-nn_plot} the results obtained in the same two mid-plane slices of the computational domain as in Fig. \ref{fig:prom-nn_red_mesh}. These results are also 
in good agreement with their HDM counterparts, which is again confirmed by the relative errors of the order of $1\%$ reported in Table \ref{table:quantitative_performance}.
More importantly however, the comparison of the results displayed in Fig.  \ref{fig:hprom-nn_plot} and Fig. \ref{fig:prom-nn_red_mesh} highlights the superior performance of the HPROM-ANN at
minimizing spurious oscillations around the traveling shock.

\newpage
\subsection{Summary and performance comparisons}

Table \ref{table:quantitative_performance} summarizes the various model parameters considered in this paper and their online performance on a single core of a Linux cluster.
It reveals that ECSW is even more computationally efficient when applied to a PROM-ANN, than it is already known to be for a traditional PROM \cite{farhat2014dimensional}.  
The results summarized in this table also show that the HPROM-ANN can deliver the same desirable level of accuracy as a traditional HPROM, using however an order of magnitude smaller dimension 
-- and consequently, is almost 6 times faster. Most importantly, the results reported in Table \ref{table:quantitative_performance} demonstrate that the HPROM-ANN can reproduce a shock-dominated 
HDM solution with 99\% accuracy, using however a four order of magnitude smaller dimension, and 50 times faster. 

\begin{table}[!h]
	\centering
	\begin{tabular}{c|c|c|c|c|c}
		Model & Dimension & $n_e$ (or $N_e$ for HDM) & $\mathbb{RE}$ & Wall clock time (seconds) & Speedup factor \\ \hline
		HDM & $125\,000$ & $62\,500$ & $-$ & $718 $ & $-$ \\ \hline
		HPROM & $95$ & $5\,689$ & $1.38\%$ & $80.5 $ &  8.9 \\ \hline
		HPROM-ANN & $10$ & $1\,496$ & $1.44\%$ & $13.8 $ & 51.9
	\end{tabular}
	\caption{Model parameters and online performance results on a single core of a Linux cluster.}
	\label{table:quantitative_performance}
\end{table}

\newpage
\section{Conclusions}
\label{sec:CONC}

The concept of a projection-based reduced-order model (PROM) augmented by an artificial neural network (ANN) -- or a PROM-ANN surrogate model -- we propose in this paper distinguishes itself 
from many previous attempts at combining PROMs and ANNs in two major aspects: 1) the size of its neural network does not scale with the large dimension $N$ of the underlying high-fidelity model (HDM), 
but with a much lower dimension related to the dimension $n \ll N$ of the PROM; and 2) it is hyperreducible to an HPROM-ANN using readily available hyperreduction methods. Consequently, the concept of an 
HPROM-ANN proposed in this paper is computationally tractable, including for very-large scale CFD models. More importantly, for a parametric shock-dominated benchmark problem known to exhibit the 
Kolmogorov $n$-width issue, our proposed HPROM-ANN delivers the same desirable level of accuracy as a traditional HPROM, using however an order of magnitude smaller dimension -- and as a result, 
operates 6 times faster. It reproduces the HDM solution with 99\% accuracy, using however a four-order-of-magnitude smaller dimension, and 50 times faster. Hence, our proposed concept of an HPROM-ANN 
has the potential for mitigating the Kolmogorov reducibility barrier. We are currently working on automating the selection of its low dimensions $n$ and $\bar n$, and confirming as well
as quantifying its potential for industry-relevant, convection-dominated, parametric, turbulent flow problems.

\section{Acknowledgments}

The authors acknowledge the support by the Air Force Office of Scientific Research under Grant FA9550-20-1-0358 and Grant FA9550-22-1-0004.

\bibliography{refs}

\end{document}